\documentclass[12pt]{article}
\usepackage{graphicx}
\usepackage{amssymb}
\input epsf

\def\beq{\begin{equation}}
\def\eeq{\end{equation}}

\title{\bf Flow coefficients in O-O Al-Al and Cu-Cu collisions at 200 GeV in the fusing color string model.}
\author{M.A. Braun$^a$, C. Pajares$^b$\\
$^a$ Dep. of High Energy physics,\\
 Saint-Petersburg State University, Russia\\
$^b$ Dep. of Particles and Instituto Gallego de Altas Energias,\\ University of Santiago de Compostela, Spain}

\begin{document}
\maketitle
\begin{abstract}
In view of the planning experiments for  collisions of light nuclei at RHIC
the   flow coefficients  for O-O, Al-AL and Cu-Cu collisions are studied
in the color string percolation model. Our results for $v_2$ are somewhat smaller
than predicted by other groups although with the same dependence on centrality.
Our obtained $v_3$ lie between predictions of other groups.

\end{abstract}

%%%%%%%%%%%%%%%%%%%%%%%%%%%%%%%%
\section{Introduction}
A remarkable discovery at collider experiments has been  observation of
strong azimuthal correlations in heavy nuclei collisions
~\cite{ref1,ref2,ref3,alver1,adamczyk,ref7}.
 It can be characterized by the non-zero
flow coefficients $v_n$ governing the correlation function of the azimuthal distribution of
secondaries as
\beq
C(\phi)=A(1+\Big(1+2\sum_{n=1}v_n\cos(n\phi)\Big).
\label{eq0}
\eeq
This phenomenon  can be interpreted in terms of formation of a fireball in the overlap of the colliding nuclei
consisting of the strongly interacting hot quark-gluon plasma, which subsequently freezes, hadronizes
and passes into the observed secondary hadrons. The dynamics of this transition seems to be well described in
the hydrodynamical approach, which relates the final spatial anisotropy to that of the initial state.

Later a similar anisotropy was found also in collisions of smaller systems such as p-p, p-A, d-A and He-A.
~\cite{khachatryan,Chatrchyan,abelev,aad,adare,adare1,aidala}.
This has raised certain doubts about formation of a significantly big pieces of quark-gluon plasma in the interaction region
and the subsequent hydrodynamical evolution. However calculations made within specific models of the latter ~\cite{bozek,sonic,iebe}
and also with  initial conditions created by  gluon emission in the Color Glass Condensate effective theory ~\cite{mace,mace1,schenke}
seem to describe at least part of the experimental data quite satisfactorily. So the dynamic assumptions adopted for A-A collisions
seem to work also for smaller systems.

This conclusion was earlier observed in an alternative scenario for high-energy collisions, namely, the fusing color string picture.
Much simpler that the hydrodynamical approach with or without previous gluon emission in the QCD framework, it allowed to
describe in a satisfactory way the dependence of the spectra both on the transverse momentum and angle at various energies and for
various colliding particles ~\cite{review}. In particular it has been found that the fusion string model correctly describes
the data on $v_n$  in pp and AA collisions ~\cite{BPV1,BP,JDD}.
In this scenario the dynamics for  small and big participants is qualitatively the same.
The colliding nucleons form strings as soon as they are close enough and the strings then emit the observed secondary particles.
The angular anisotropy in this scenario is the result of their quenching due to the presence of the gluon field  from the created strings.
So essentially it is a two-stage scenario as opposed to three-stage scenarios consisting first in formation of the set of interacting nucleons,
then building of the initial condition  ( e.g. emission of gluons)  and finally the hydrodynamical expansion. Correspondingly it carries
only one adjustable parameter - the universal quenching coefficient to be extracted  from some data.

Remarkably this approach does not distinguish between colliding particles of different hadronic content. In particular it has been found that
it well applies not only to heavy nuclei collisions but also to p-A and d-A collisions ~\cite{BP1}.
Accordingly in this note we apply this approach to the flow coefficients $v_2$. $v_3$  and $v_4$ in O-O, Al-Al  and Cu-Cu collisions at 200 GeV now planned at RHIC.
We expect that our results will serve as one of reference points for the future experimental data. Our hope is that they will describe them reasonably well.

The flow coefficients in O-O collisions have recently be studied in the approaches with either hydrodynamical evolution from the initial condition created in the color gluon condensate
(CGC) framework ~\cite{schenke1, schenke2} or with the initial parton production and the following rescattering and hadronization (AMTP approach, see e.g. ~\cite{chinese}.
Both models are much more complicated than our color string model. They distinguish the initial state formation and final state interaction. And as stated in ~\cite{schenke}
the question which of the two stages dominates in producing the final azimuthal dependence remains open. As mentioned the color string scenario is basically much simpler than both.
In this approach particle production and their azimuthal asymmetry are produced simultaneously due to asymmetry of the created gluon field.
As we shall see it gives the flow coefficients for light nuclei on the same level and with the same centrality dependence as the two mentioned
sophisticated approaches. So it seems that the complicated concrete mechanisms inherent in the latter are in fact not very essential for the  final results, which apparently depend
only on the overall basic dynamical structure, correctly described by the color string model

\section{Flow coefficients in the color string model}
.

The model was proposed a long time ago to describe
multiparticle production in the soft region. Its latest development and applications are described in the review paper ~\cite {review}.
As mentioned in this model particle creation and production of azimuthal asymmetry proceed simultaneously.
Both are the results of the formation of the gluon field by color strings. This field, on one hand, produces particles more or less
in the spirit of the Schwinger mechanism of particle creation in the external field. On the other hand, it provides the
azimuthal asymmetry due to interaction of the produced particles with the same gluon field, which by itself is azimuthal asymmetric
due to the initial asymmetry of nucleons in the overlap and fluctuations.

In the model the initial strings tend to overlap and fuse into novel strings with more intrinsic color which have a higher tension and
so partons with greater transverse momenta.
As atomic numbers of the colliding nuclei grow the density of strings in the transverse space grows. As a result, with the growth of their density
strings fuse more intensely and this effective number grows weaker. This immediately explains the observed growth of the multiplicities in AA collisions
both with the atomic numbers and energy. Also appearance of fused strings explains the growth of the heavy particles in these collisions. At a certain
critical density clusters acquire the transverse dimensions
comparable to that of the
interaction area (percolation) and one may consider it as formation of the drops of the quark-gluon plasma.

Naively one can assume that strings
decay into particles ($q\bar{q}$ pairs) by the well-known  Schwinger mechanism for
pair creation in a strong
electromagnetic field.
\beq
P(p,\phi)=Ce^{-\frac{p_0^2}{T}},
\label{prob}
\eeq
where $p_0$ is the particle initial transverse momentum,
$T$ is the string tension (up to an irrelevant numerical coefficient) and
$C$ is the normalization factor. The string tension is determined by the magnitude of the gluon field responsible for particle creation.
To extend  validity of the distribution to
higher momenta one may use the idea that the field and consequently string tension fluctuate, which
transforms the Gaussian distribution into the thermal one ~\cite{bialas,deupaj}:
\beq
P(p,\phi)=Ce^{-\frac{p_0}{t}}
\label{probb}
\eeq
with temperature $t=\sqrt{T/2}$.

The initial transverse momentum $p_0$ is thought to be
different from the
observed particle momentum $p$ due to particle interaction with the same gluon field.
In fact the particle has to pass through the  fused string areas and emit gluons
on its way out. So in Eq. (\ref{prob}) or (\ref{probb}) one has to consider $p_0$
as a function of $p$ and path length $l$ inside each string encountered on its way out:
$p_0=f(p,l(\phi))$ where $\phi$ is the azimuthal angle. It is this quenching that creates the final anisotropy
and leads to nonzero flow coefficients, due to anisotropy of string distribution.
To describe this quenching
we use the corresponding QED picture for a charged particle  moving
in the external electromagnetic field ~\cite{nikishov}.
This leads to
the quenching formula inside a string passed by the parton~\cite{ref14}
\beq
p_0(p,l)=p\Big(1+\kappa p^{-1/3}T^{2/3}l\Big)^3,
\label{quench1}
\eeq
Here $l$ is the length traveled by the parton through the gluon fields in the hadron
formed by color strings. Note that both $l$ and $T$ are different for different strings.
When the parton passes through many strings inside the hadron one should
sum different $T^{2/3}_1l_1+ T^{2/3}_2l_2+...$ over all of them.
For an event both $T_i$ and $l_i$ are uniquely determined by the geometry of the collision
and string fusion. The quenching coefficient $\kappa$ to be taken from the experimental
data. We adjusted $\kappa$  to
give the experimental value for the coefficient $v_2$
in mid-central Pb-Pb collisions at 5-13 TeV GeV, integrated over the
transverse momenta, which gives $\kappa=0.45$.

Remarkably, Eq. (\ref{quench1}) gives rise to a universal dependence of $v_2$ on the
product  $\epsilon p^{2/3}T^{1/3}l$, where $\epsilon$ is the eccentricity of the nuclear overlap.
This scaling is well confirmed by the experimental data ~\cite{andres1,andres2,acharva}.

In the c model the event is realized by a particular way of exchange of color strings
between the projectile and target. Different events possess  different number of strings located at different places
in the overlap of the colliding nuclei.
The flow coefficients are obtained after averaging over events of the inclusive particle distribution
in the azimuthal angle for a single event
\beq
I^e(\phi)=
A^e\Big[1+2\sum_{n=1}\Big(a_n^e\cos n\phi+
b_n^e\sin n\phi\Big)\Big].
\label{ie}
\eeq
The flow coefficients are
\beq
v_n=\Big<\Big[(a_n^e)^2+(b_n^e)^2\Big]^{1/2}\Big>.
\label{vnexp1}
\eeq

In experimental observations one often uses instead of (\ref{vnexp1})
\beq
v_n\{2\}=\Big(\Big<(a_n^e)^2+(b_n^e)^2\Big>\Big)^{1/2}.
\label{vnexp2}
\eeq
which is somewhat greater than $v_n$ defined by (\ref{vnexp1})

\section{Calculations}
\subsection{Monte-Carlo simulation}

The Monte-Carlo simulation with the string model in principle consist, first, of distributing strings in the transverse space of the overlapping nuclei
second in analyzing their geometrical structure and forming fused strings when they overlap and third, studying the paths passed by the produced particles
through the maze of strings to find their quenching according to Eq (\ref{quench1}).
The actual realization of this program encounters with the fact that
 the fusing strings
form clusters of different forms and dimension. Both the study of their emission and their quenching of already emitted partons
presents practically unsurmountable technical problems if ones considers their exact geometries. To facilitate the problem
we use a simplified  approach which was shown to give a very reasonable result as compared to the exact one ~\cite{BKPV}.
Instead of clusters formed by the actual fusing of formed strings we split the whole transverse area
in cells having the dimension of ordinary strings ($\sim 0.3 $ fm). Distributing the strings in the area we
consider as fused the ones which get into the same cell. Then the geometry is simplified to the set of clusters represented by cells which contain fused
strings made of different number of ordinary ones. Each cell-cluster possesses its own string tension and emission multiplicity in accordance with the
standard fusing string model. The quenching of the emitted parton is then studied in this geometry as this parton passes through different cells on his way out.

Some problem is also that in the Monte-carlo simulation one has to place the colliding particles at a fixed value of the impact parameter $b$.
However in the experimental setup the impact parameter is not known.
Instead one can distinguish between different centralities determined from the observed multiplicities. To find observables as function of the multiplicity
one has first to study the multiplicity at different values of $b$ and then divide it into the intervals corresponding to centralities as defined by the experimentalists
and study separately Monte-Carlo simulations within a given interval of multiplicities. This does not seem to present much difficulty but to have a
reasonable statistics for each interval of centrality it leads to considerable rise in the total number od simulations and thus to the rise of the
total simulation time.

\subsection{Results}
Calculations were performed for O-O , Al-Al and Cu-Cu collisions at 200 GeV/c. The reported run consists of 400 simulations divided between 11 centrality
intervals from $0\%$ to $100\%$. The interval of transverse momenta was taken as $0.1<p_T<4$ GeV/c divided in 80 points.
The interval of azimuthal angles was divided in 360 points.  Our flow coefficients were calculated according to Eq. (\ref{vnexp2}).

\begin{figure}
\begin{center}
\includegraphics[width=7.2 cm]{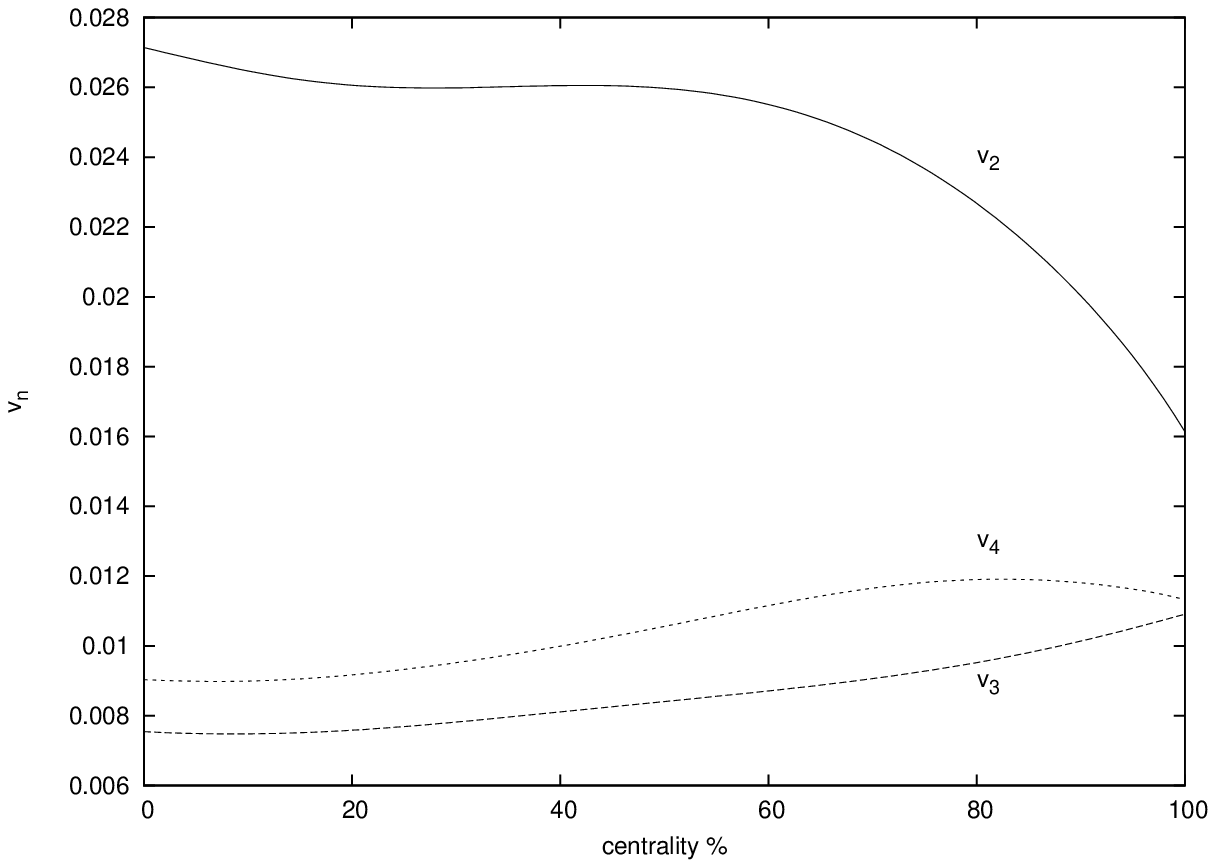}
\includegraphics[width=7.2 cm]{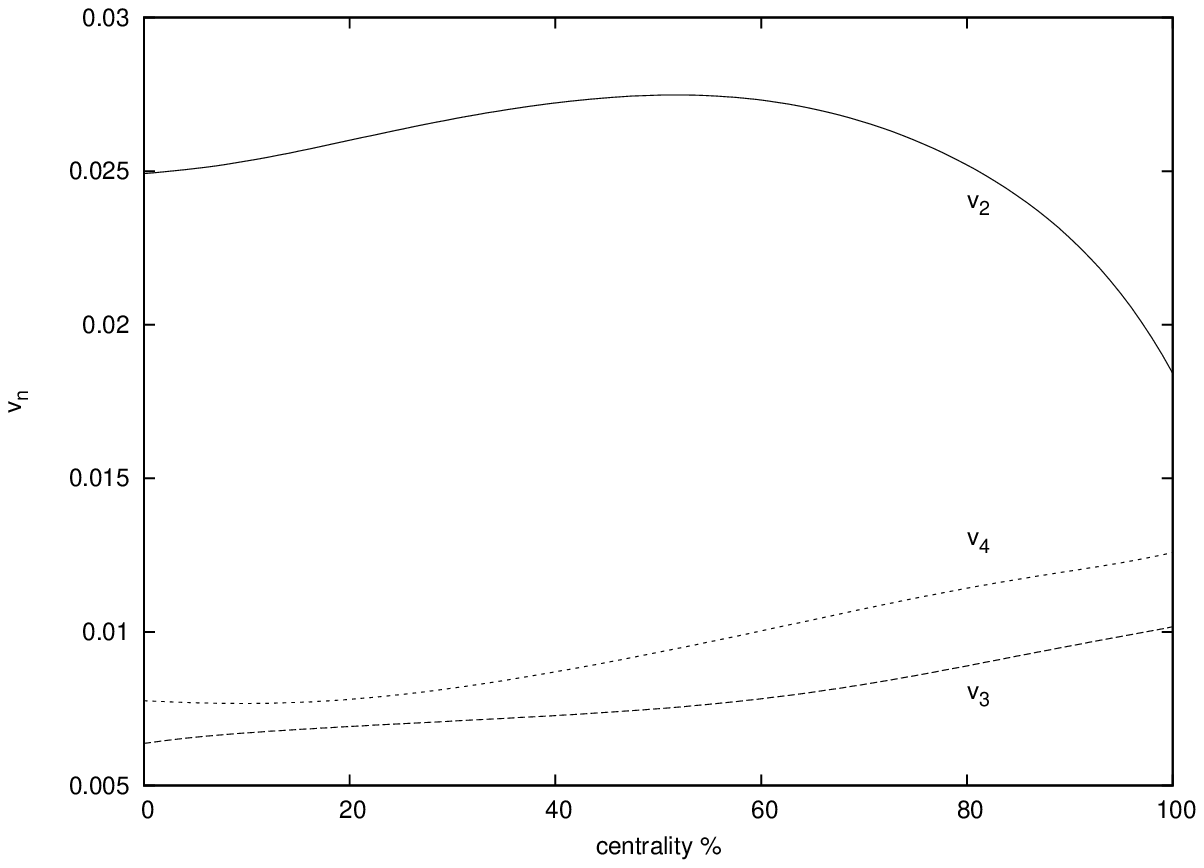}
\includegraphics[width=7.2 cm]{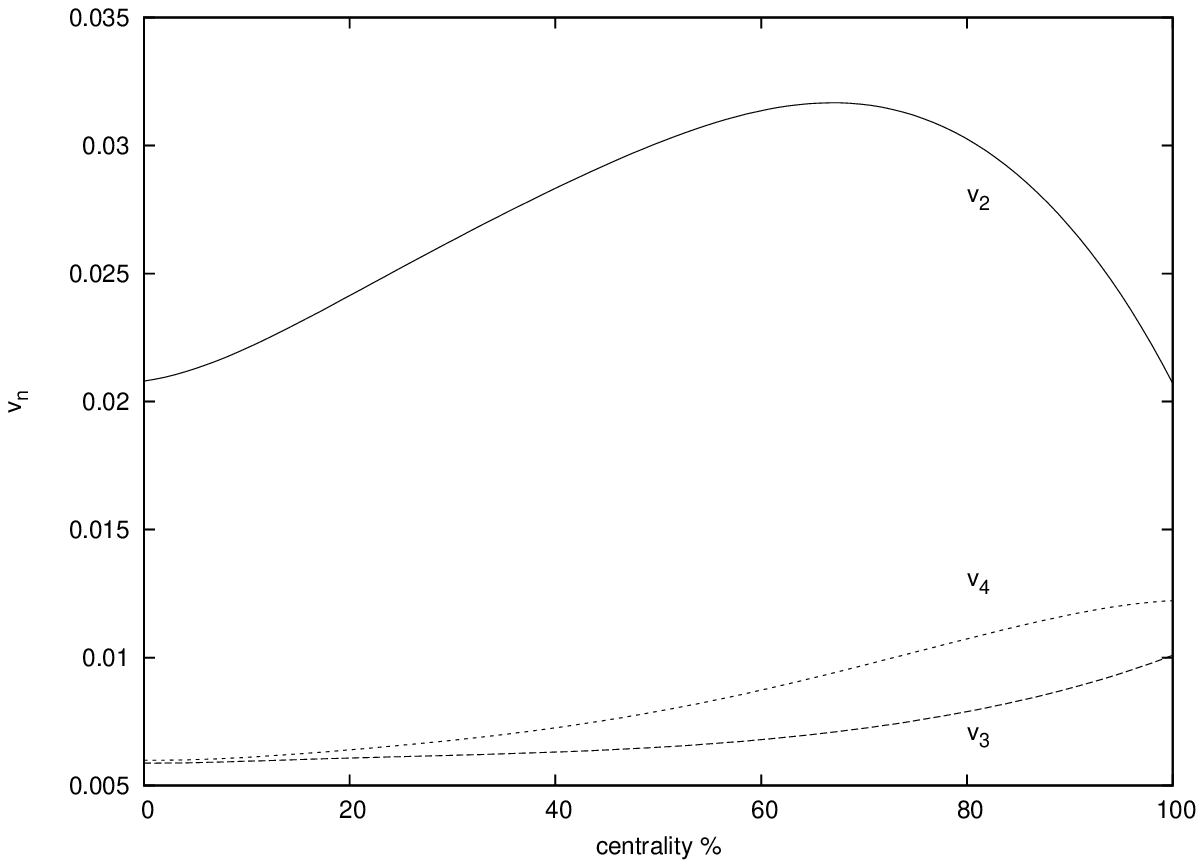}
\caption{The calculated flow coefficients $v_2$, $v_3$ and $v_4$  for O-O (left upper panel) Al-Al (right upper panel)
and Cu-Cu (lower panel) collisions at 200 GeV/c as function of centrality}
\label{fig1}
\end{center}
\end{figure}

The results of  calculations for $v_2$, $v_3$ and $v_4$ for different centralities are presented in Fig. \ref{fig1}
for O-O, Al-Al and Cu-Cu collisions.

In the next figure we illustrate the dependence of $v_2$, $v_3$ an $v_4$  on the atomic number1 16,27 and 64 for
for O-O, Al-Al and Cu-Cu collisions.
\begin{figure}
\begin{center}
\includegraphics[width=7.2 cm]{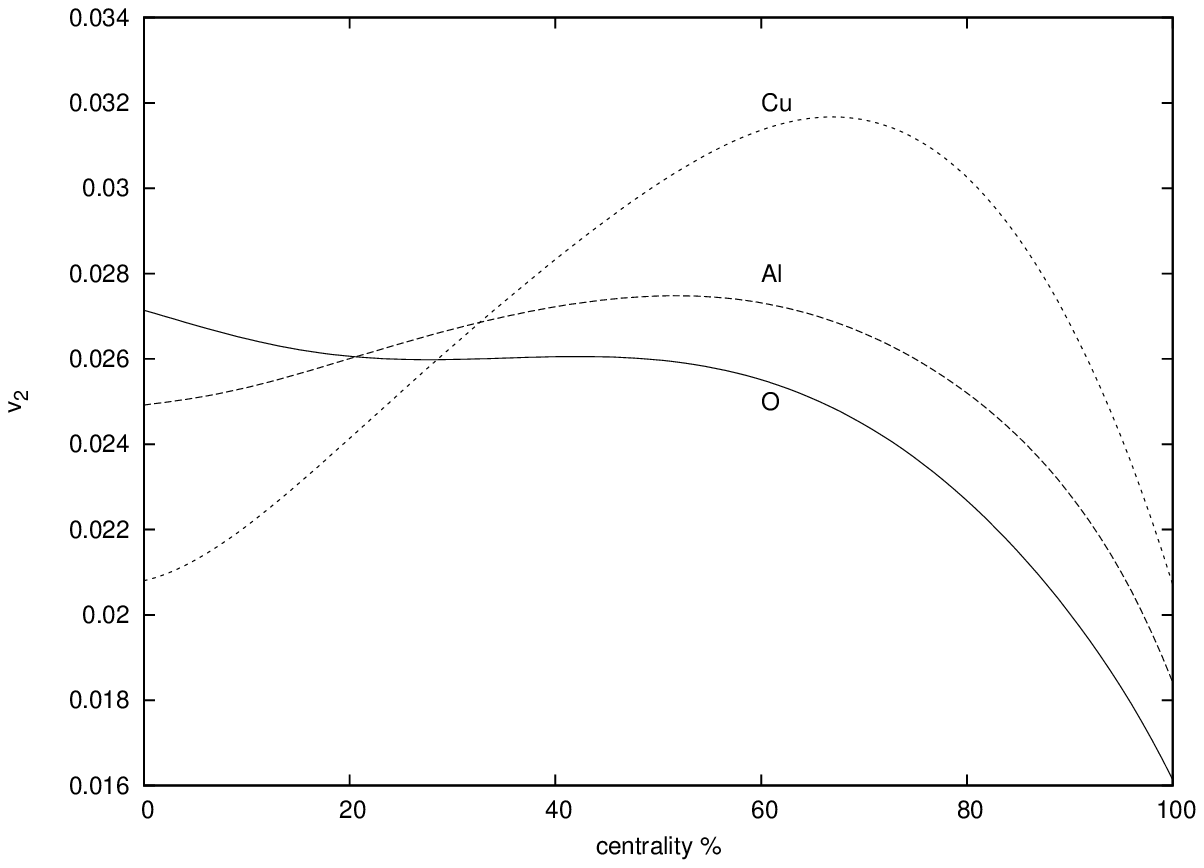}
\includegraphics[width=7.2 cm]{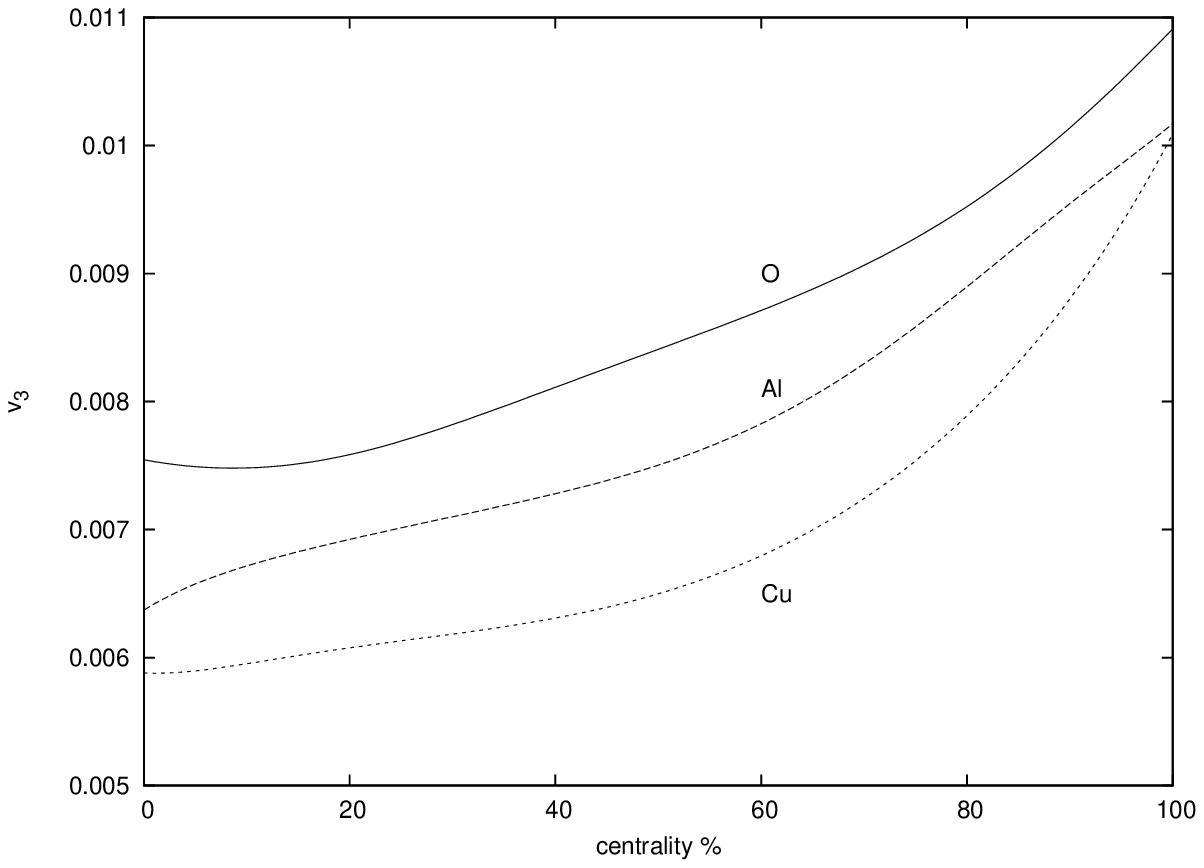}
\includegraphics[width=7.2 cm]{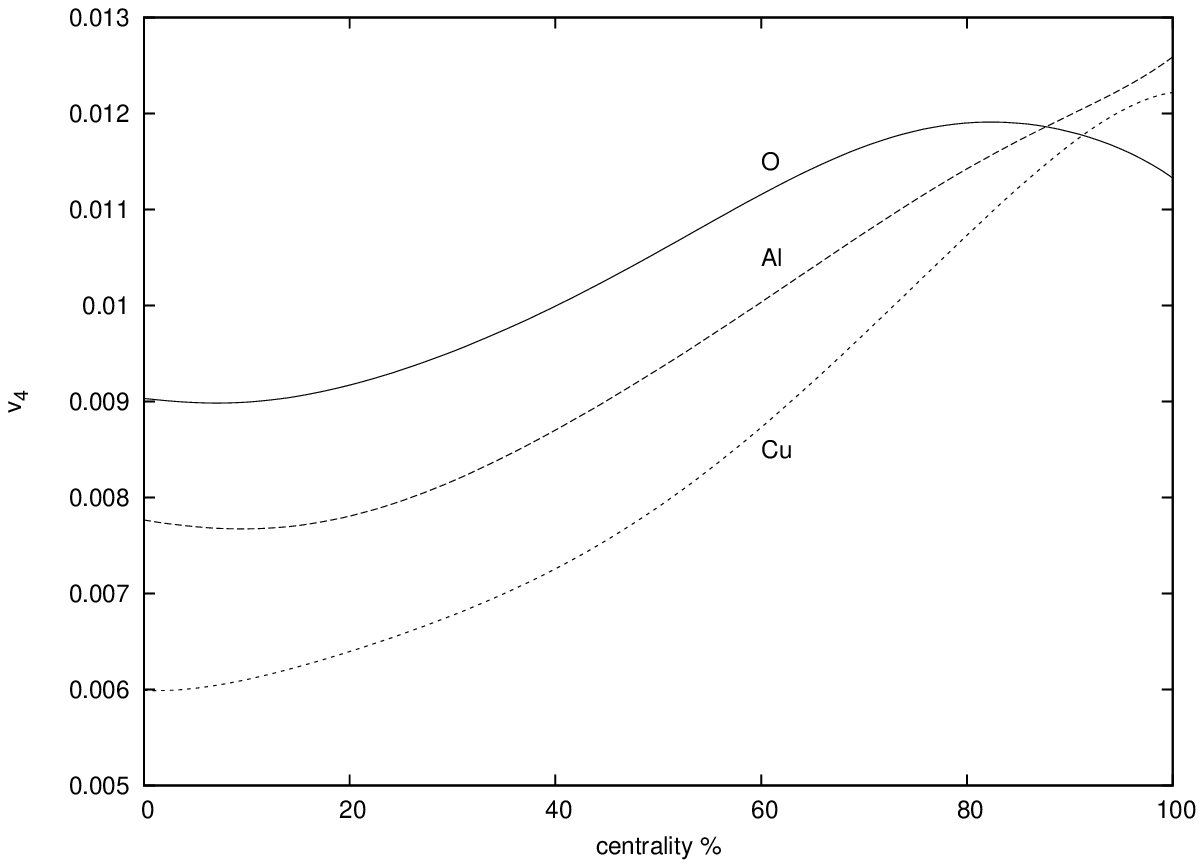}
\caption{The calculated flow coefficients $v_2$  (left upper panel) $v_3$ (right upper panel) and $v_4$ (lower panel)
for O-O, Al-Al and Cu-Cu collisions at 200 GeV/c as function of centrality}
\label{fig2}
\end{center}
\end{figure}

\begin{figure}
\begin{center}
\includegraphics[width=7.2 cm]{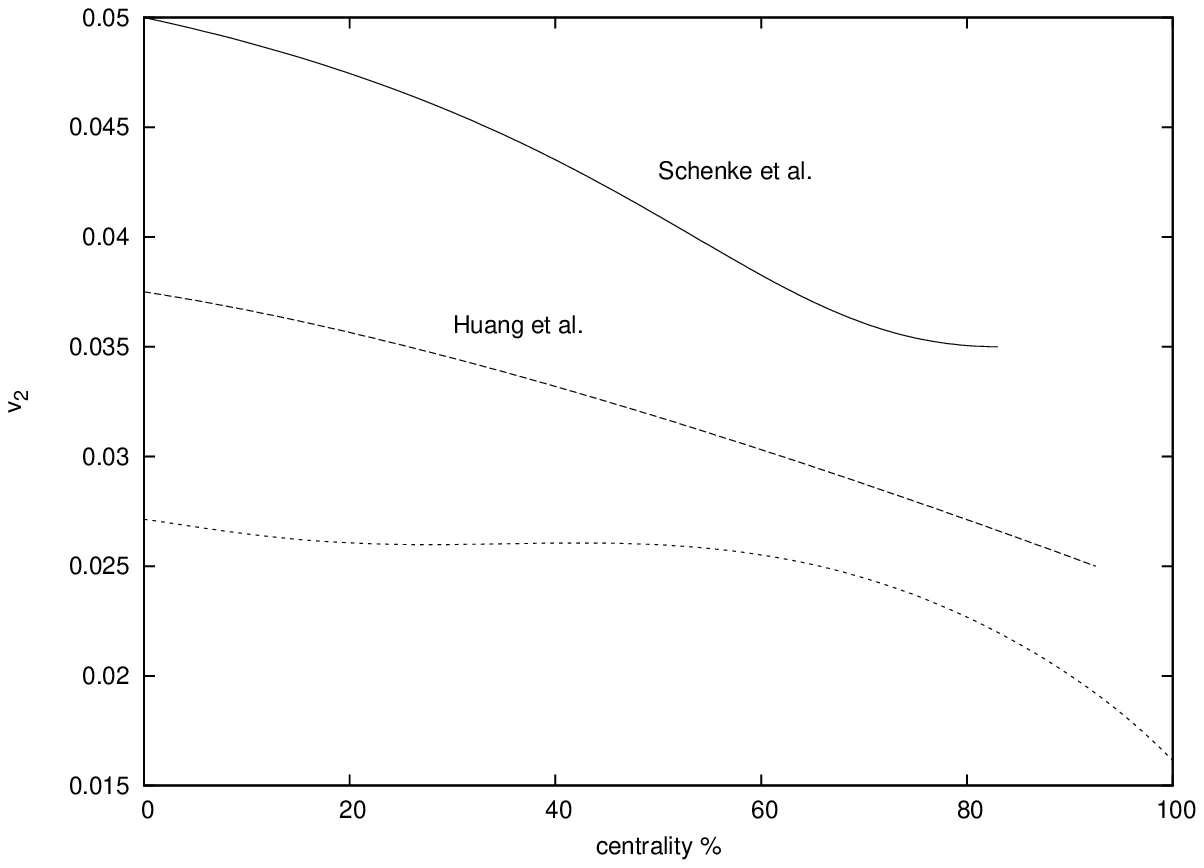}
\includegraphics[width=7.2 cm]{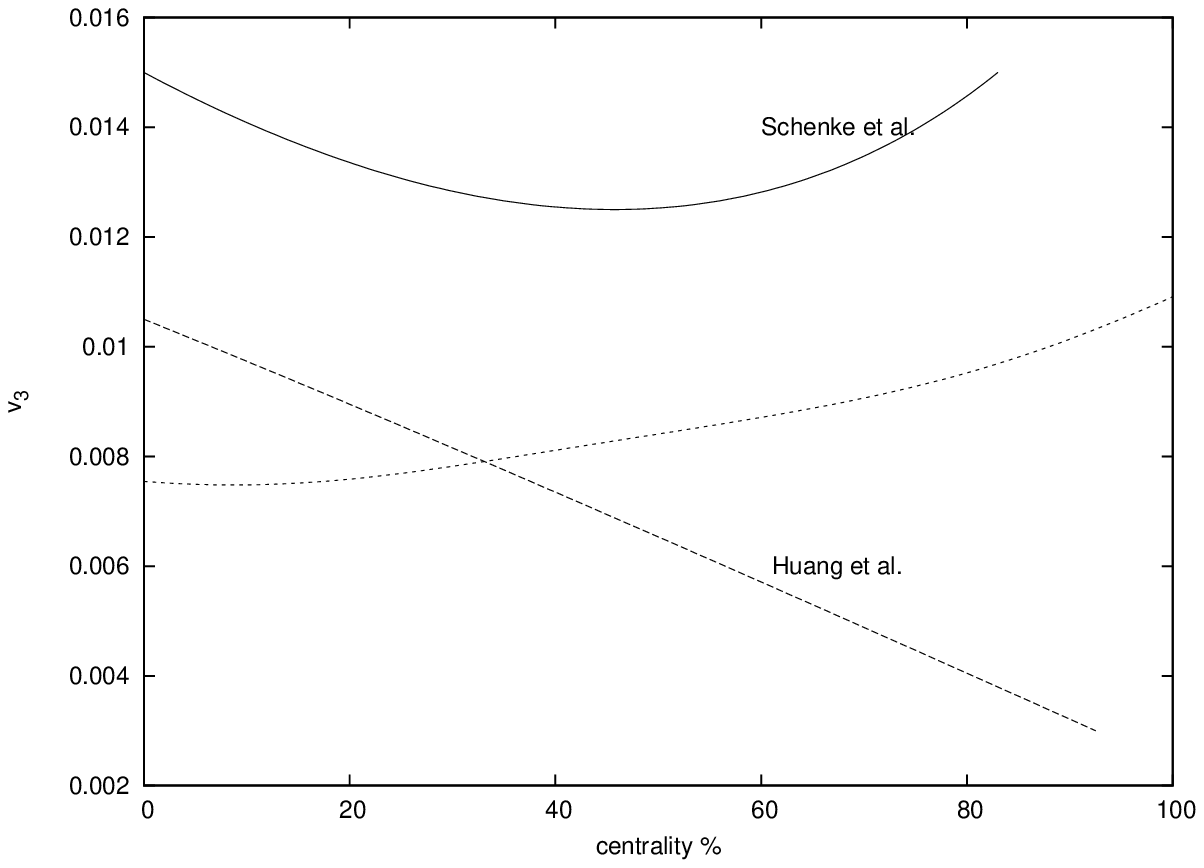}
\caption{The calculated flow coefficients $v_2$ (left panel) and $v_3$ (right panel)  for O-O collisions at 200 GeV/c
compared to  ~\cite{schenke2} and~\cite{chinese}}
\label{fig3}
\end{center}
\end{figure}

In Fig. \ref{fig3} we compare our calculated $v_3$ and $v_3$ with the predictions of ~\cite{schenke} and~\cite{chinese}.

Finally we illustrate the $p_T$ -dependence of $v_n$ in 0-0 collisions at different centralities. The centrality interval in each case lies within $5\%$
around the indicated in the figures. In this case $v_3$ turns out invariably greater than $v_4$. Since $v_n$ at different centralities
are not averaged $v_n(p_T)$ this seems to be not in contradiction with Fig. \ref{fig1}.

\begin{figure}
\begin{center}
\includegraphics[width=7.2 cm]{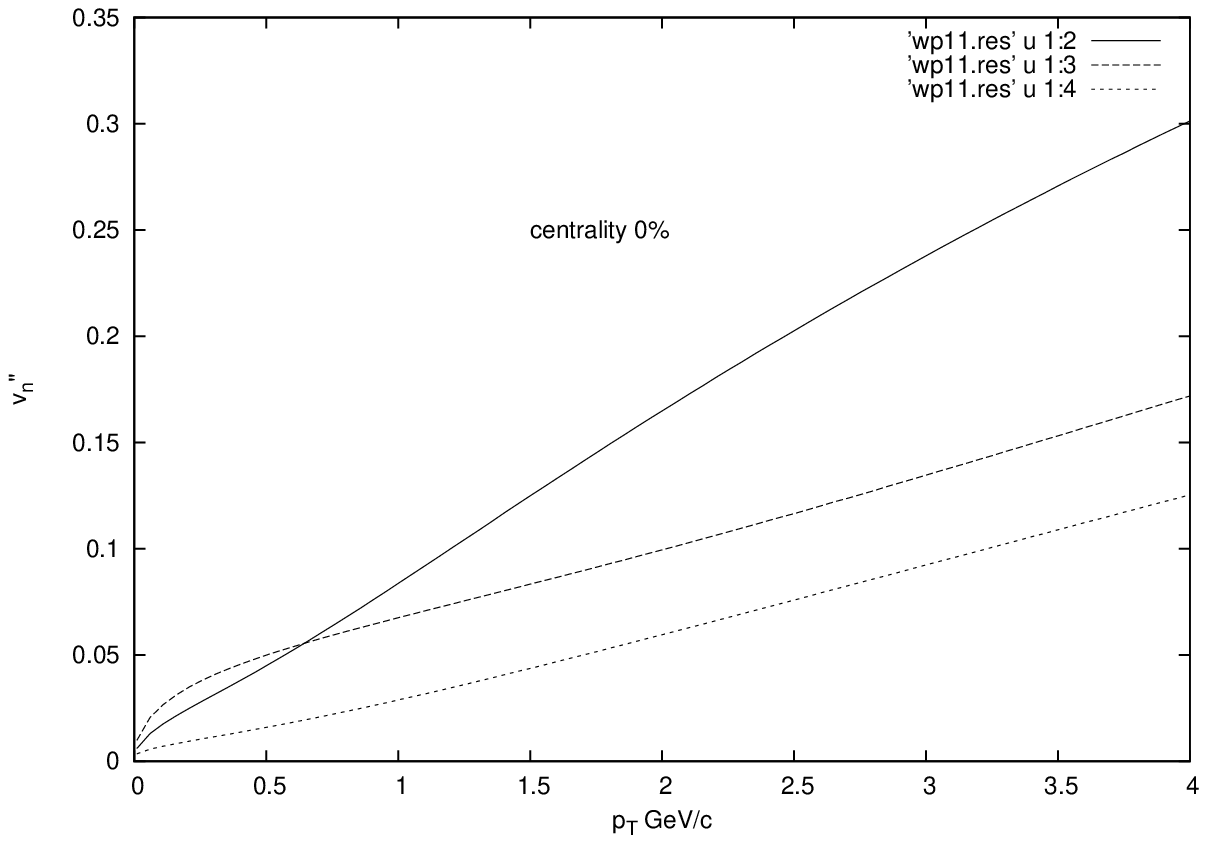}
\includegraphics[width=7.2 cm]{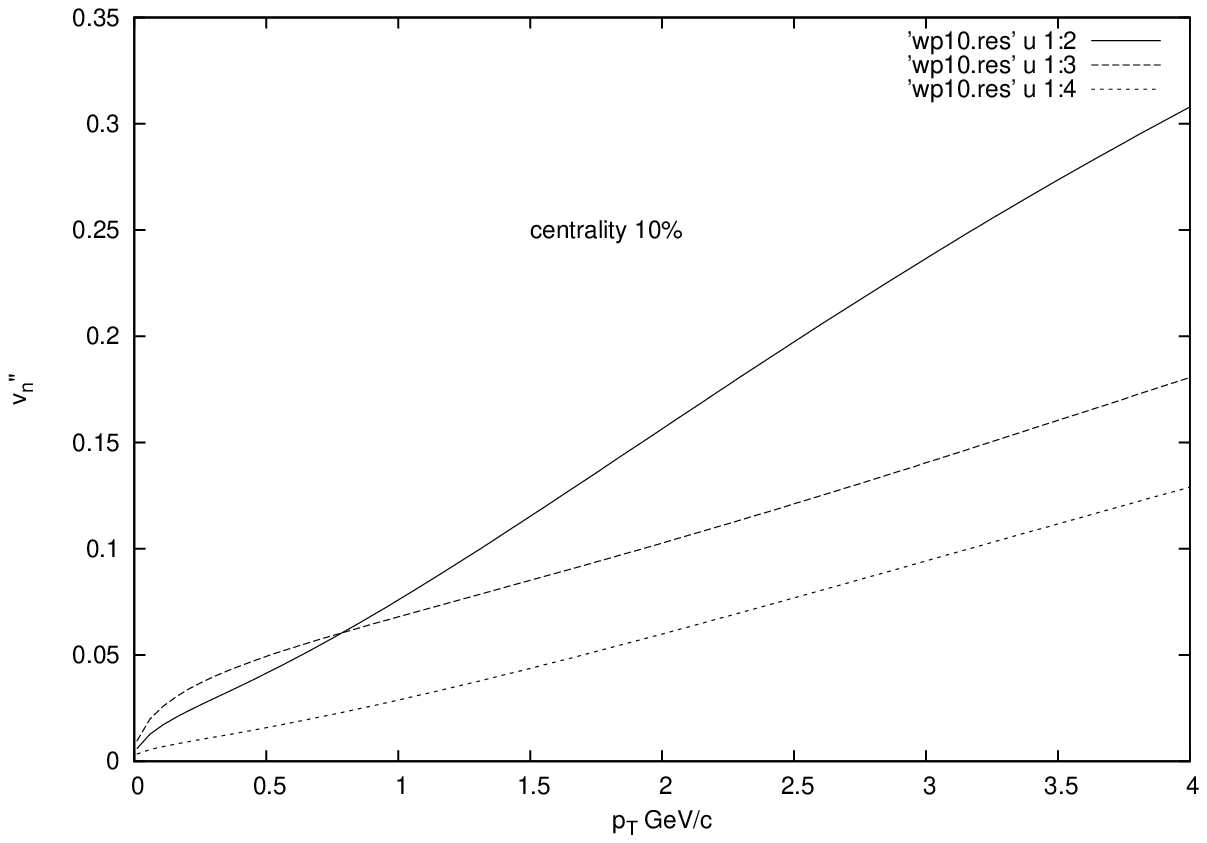}
\includegraphics[width=7.2 cm]{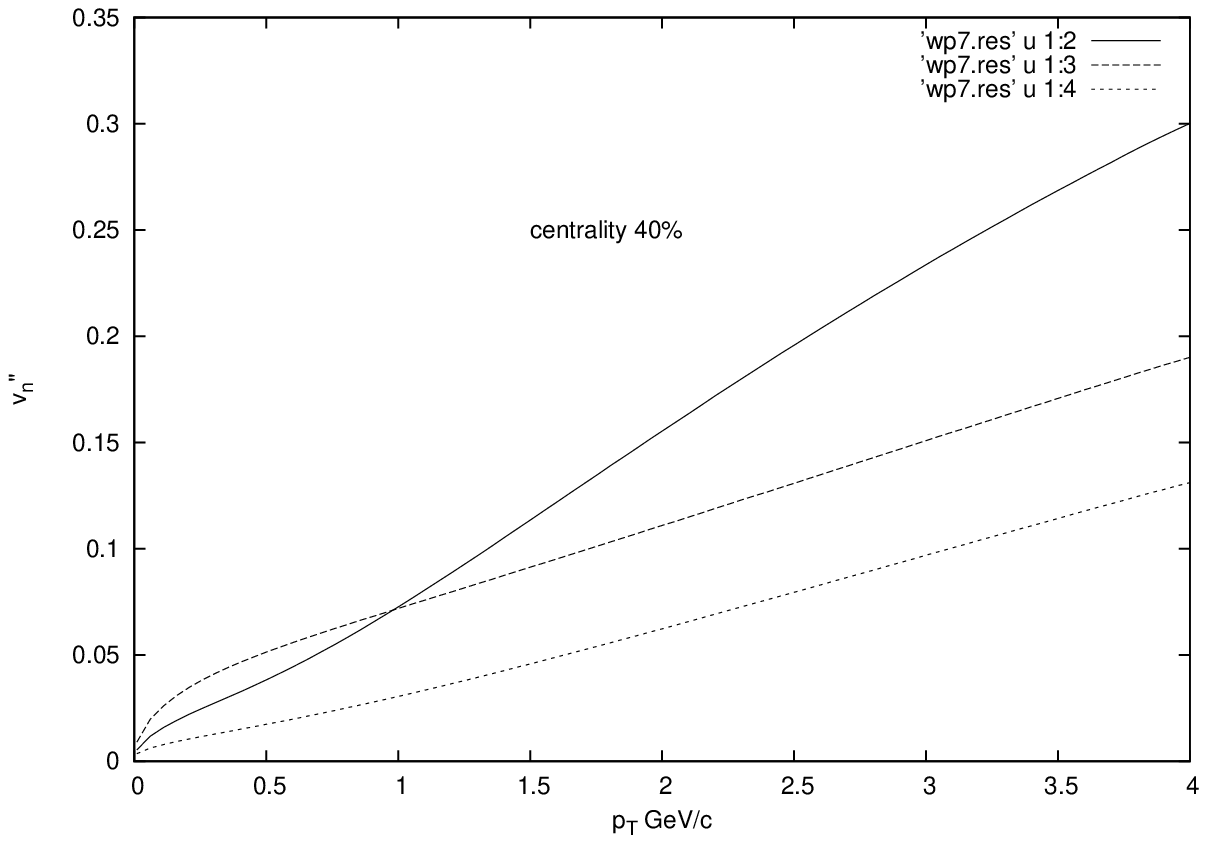}
\includegraphics[width=7.2 cm]{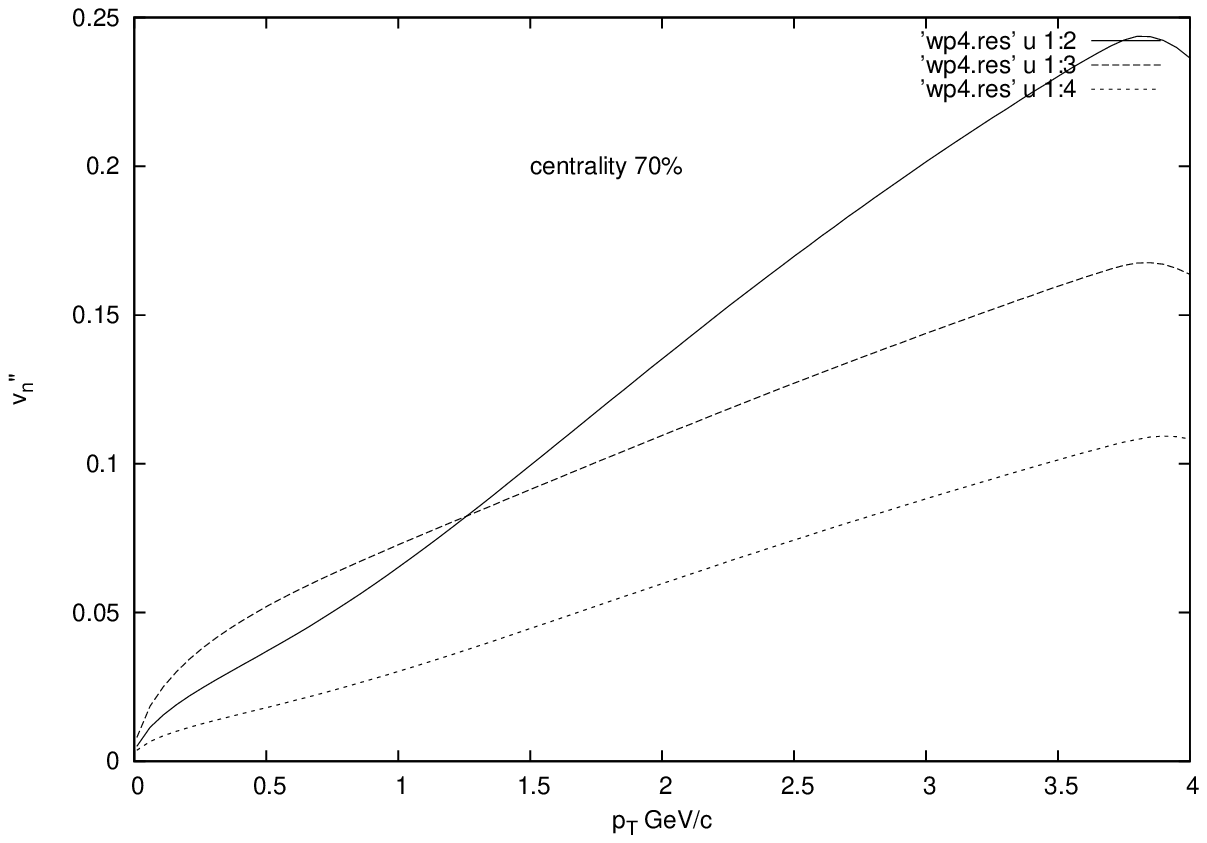}
\caption{The calculated flow coefficients $v_2$, $v_3$ and $v_4$  for O-O collisions at 200 GeV/c as function of $p_T$
at centralities $0\div 5$ (left upper panel) $10\div 15$ (right upper panel) $40\div 45$ (left low panel) and
$70\div 75$ (right lower panel)}
\label{fig9}
\end{center}
\end{figure}

One has to take into account that these results were obtained by Monte-Carlo simulations with a number of runs limited by
our calculational possibilities to $N_{run}=400$. According to the general properties of the Monte Carlo procedures this gives a rough
estimate of our precision $\sim 1/\sqrt{N_{run}}\sim 5\%$.

\section{Discussion}
Our predictions for the elliptic flow  $v_2$  are roughly of the same magnitude for O-O, Al-Al and Cu-Cu collisions at 200 GeV with a rather weak centrality dependence.
Both triangular and quadrangular flows $v_3$ and $v_4$ turn out to fall with A and rise with centrality.
Remarkably, as a function of centrality,  the quadrangular flow is found to be larger than the triangular one, although it is smaller as a function of $p_T$.
Compared to the calculations of other groups,
our $v_2$ are definitely smaller than in ~\cite{schenke2} (by nearly 40\%) and ~\cite{chinese} (by $\sim$30 \%) with a similar centrality dependence.
As to $v_3$  our predictions are also smaller than in ~\cite{schenke2} with the same dependence on centrality. Compared to ~\cite{chinese} our $v_3$ behave differently in centrality.
lower at small and higher at large centralities. One may say that on the average they are of the same magnitude.
We stress that the azimuthal asymmetry in our model is directly controlled by the quenching parameter $\kappa$ in Eq. (\ref{quench1}). As mentioned, in these calculations
it was chosen 0.45 from the previous  similar studies of Pb-Pb collisions at LHC. So it is trivial to somewhat raise our flows by increasing parameter $\kappa$. However it would contradict
our physical picture in which $\kappa$ is determined by the quenching of a particle passing through the gluon  field independent of such external parameters as the atomic number of
participants and their energy.

\section{Acknowledgements}
M.A.B. appreciates hospitality and financial support of the University of Santiago de Compostela, Spain.
C.P. thanks the grant Maria de Maeztu Unit of Excelence of Spain and the support
Xunta de Galicia and FEDER.
. This work was partially done under the project EPA 2017-83814-P
of Ministerio Ciencia, Tecnologia y Universidades of Spain.

\end{document}